\documentclass[prl,aps,floatfix,amsmath,amssymb,superscriptaddress,tightenlines,twocolumn]{revtex4}
\usepackage{graphicx}
\usepackage{epstopdf}
\newcommand{\be}{\begin{equation}}
\newcommand{\ee}{\end{equation}}
\usepackage{amsmath}
\usepackage{amsfonts}
\usepackage{bm}
\usepackage{color}
\usepackage{subfigure}
\usepackage{amsthm}
\usepackage[latin1]{inputenc}
\usepackage{tikz}
\usepackage{comment}

\newcommand{\Tr}{\textrm{Tr}}

\newtheorem{thm}{Theorem}

\begin{document}
\title{On the informational completeness of local observables}
\author{Isaac H. Kim}
\affiliation{Perimeter Institute of Theoretical Physics, Waterloo ON N2L 2Y5, Canada}
\affiliation{Institute for Quantum Computing, University of Waterloo, Waterloo ON N2L 3G1, Canada}

\date{\today}
\begin{abstract}
For a general multipartite quantum state, we formulate a locally checkable condition, under which the expectation values of certain nonlocal observables are completely determined by the expectation values of some local observables. The condition is satisfied by ground states of gapped quantum many-body systems in two spatial dimensions, assuming a widely conjectured form of area law is correct. Its implications on the studies of gapped quantum many-body systems, quantum state tomography, and quantum state verification are discussed. These results are based on a partial characterization of states with small yet nonzero conditional mutual information, which may be of independent interest.
\end{abstract}

\maketitle

Many difficulties in the studies of interacting quantum many-body systems are often attributed to the exponential blowup of the underlying Hilbert space dimension. As such, it is desirable to identify a special structure that can be justified from a set of physical principles, and use the structure to reduce the complexity of the problem.

We do just that in this manuscript. Here, the physical principle is the area law, an observation that entanglement entropy of physical states are often proportional to its boundary area.\cite{Eisert2008} Positing area law, we derive a number of highly nontrivial consequences in seemingly disparate fields, ranging from the studies of quantum many-body systems, to quantum state tomography,\cite{Vogel1989} and quantum state verification.\cite{Flammia2011,Silva2011} All of these results are rooted in the same idea: that area law implies informational completeness of local observables. Naturally, explaining this idea shall be the primary focus of this manuscript.

Let us first define what it means for a set of observables to be informationally complete. A quantum state describing $n$ qubits can be expressed as a $2^n \times 2^n$ positive semi-definite matrix with a unit trace. One can specify some of the matrix entries by assigning expectation values to some observables. If the assignment completely determines the state, we say the set of observables is informationally complete. One trivial example would be to simply assign expectation values to all the observables. On the other hand, if the number of assignments are significantly smaller than the number of matrix entries, one may expect the state to be ill-defined. 

We prove a fact that is counter to this intuition: that one can compute the expectation values of all the observables from the expectation values of strictly local observables alone, assuming the local expectation values satisfy a certain condition. If the condition is satisfied, the global state is completely determined by its local reduced density matrices; this is what we mean by the informational completeness of local observables. In order to explain the idea clearly, we shall focus on a specific family of states: states that obey area law in two spatial dimensions(2D). This encompasses practically all the known quantum many-body ground states in 2D that are separated from the excited states by a uniform energy gap. We suspect the condition we obtained in this manuscript is satisfied by other states as well, but we leave that for future work.

The fact that local observables can be informationally complete has been noted previously by Cramer et al.\cite{Cramer2011} Our result has a similar flavor, but there are also important differences. For example, Ref.\cite{Cramer2011} works for states that are well-approximated by a matrix product state, but our approach is completely independent of any variational wavefunctions. In particular, our result is not restricted by the dimensionality of the underlying lattice.

Similar to Ref.\cite{Cramer2011}, we provide a rigorous bound on the distance between two quantum states, $\rho$ and $\sigma$, that are consistent over a set of local subsystems. Denoting the set of subsystems as $\{S_i \}$, our bound has the following form:
\begin{equation}
|\rho - \sigma|_1 \leq \epsilon(\{\rho_{S_i}\}), \label{eq:Main_Result}
\end{equation}
where $| \cdots |_1$ is the trace distance and $\epsilon(\{\rho_{S_i} \})$ is a number that can be easily computed from the local reduced density matrices.

The result is explained in two steps. The first step provides an upper bound on the trace distance between two quantum states that are locally indistinguishable from each other. The upper bound involves the global state, which motivates the second step; we prove a weaker upper bound which has the virtue of involving only the local reduced density matrices. 

Once we obtain the final form of Eq.\ref{eq:Main_Result}, we shall apply it to systems that obey area law. The informational completeness of local observables for such systems will follow immediately. Its implications on the studies of quantum many-body systems, quantum state tomography, and quantum state verification will be also explained.

{\it Approximately conditionally independent states---} We begin by proving a property of approximately conditionally independent states, which is new to the best of author's knowledge. 

Let us first define the relevant notations. For a tripartite quantum system, a state is conditionally independent if the conditional mutual information is equal to $0$. Denoting each of the subsystems as $A,B,$ and $C$, the conditional mutual information is defined as follows:
\begin{equation}
I(A:C|B) \equiv S(A|B) - S(A|BC) \geq 0, \nonumber
\end{equation} 
where $S(A|B) = S(AB) - S(B)$ is the conditional entropy. The inequality is the strong subadditivity of entropy, a statement that holds for any quantum states.\cite{Lieb1972} A tripartite state is referred to be approximately conditionally independent if its conditional mutual information is close to $0$, as opposed to being exactly equal to $0$. If two states are equivalent over $AB$ and $BC$, they are locally equivalent;  if their global states are equivalent, they are globally equivalent.

It is known that two conditionally independent states are globally equivalent if and only if they are locally equivalent. This is based on a rather deep result in operator theory, which asserts that the following formula holds if and only if the state is conditionally independent\cite{Petz2003}:
\begin{equation}
\rho_{ABC} = \rho_{AB}^{\frac{1}{2}} \rho_B^{-\frac{1}{2}}  \rho_{BC} \rho_B^{-\frac{1}{2}} \rho_{AB}^{\frac{1}{2}}. \label{eq:Petz}
\end{equation}
For the purpose of this manuscript, the precise form of the formula is not particularly important; the important point is that the global state is completely specified by its marginal distributions. Applying Eq.\ref{eq:Petz} to both states, one can easily verify that local equivalence implies global equivalence for such states.

Our key technical result here is that the aforementioned statement is robust; see Theorem 1.
\begin{thm}
For tripartite quantum states $\rho$ and $\sigma$ over $ABC$, if $\rho_{AB}=\sigma_{AB}$ and $\rho_{BC} = \sigma_{BC}$,
\begin{equation}
\frac{1}{4}|\rho - \sigma|_1^2 \leq I(A:C|B)_{\rho} + I(A:C|B)_{\sigma},
\end{equation}
where the subscript denotes the underlying quantum state.
\end{thm}

Since we are interested in multipartite states, a multipartite generalization of Theorem 1 is certainly desirable at this point. We introduce a concept that becomes particularly handy in understanding such generalization: Markov entropy decomposition. Markov entropy decomposition is a decomposition of the global entropy into local entanglement entropies.\cite{Poulin2010a} For a quantum state $\rho$, Markov entropy is defined as follows:
\begin{equation}
S_M(\rho) \equiv \sum_{k=1}^{N} S(k | \mathcal{M}_k)_{\rho}, \nonumber
\end{equation}
where the subscript on the conditional entropy refers to the underlying quantum state, and $\mathcal{M}_k$ is the so called Markov shield, through which correlations are mediated.\cite{Poulin2010a} 

The generalization to the multipartite setting is presented below.  
\begin{thm}
If $|\rho_{k\mathcal{M}_k}- \sigma_{k\mathcal{M}_k}|_1=0$ for all $k$, 
\begin{equation}
\frac{1}{4}|\rho - \sigma|_1^2 \leq S_M(\rho) - S(\rho) + S_M(\sigma) - S(\sigma). \label{eq:MEDLowerBound}
\end{equation}
\end{thm}
If the trace distance between $\rho_{k\mathcal{M}_k}$ and $\sigma_{k\mathcal{M}_k}$ are close to $0$, as opposed to being exactly equal to $0$, Eq.\ref{eq:MEDLowerBound} is modified by an additive error term. The general proof that includes such term shall be given in the supplementary material. 

{\it Locally checkable conditions---}  As it stands, the upper bound in Theorem 2 is not terribly useful for our purpose. It involves the global state, while we explicitly stated that the upper bound in Eq.\ref{eq:Main_Result} only involves local reduced density matrices. Our resolution is to provide a locally computable upper bound to the right-hand-side of Eq.\ref{eq:MEDLowerBound}. The new upper bound is weaker, but it only involves the local reduced density matrices.

Our argument relies on a variant of SSA, which is known as the weak monotonicity(WM):
\begin{equation}
S(A|B) + S(A|C) \geq 0. \nonumber
\end{equation}
WM is equivalent to SSA, in a sense that one can be derived from another by purifying the global state.\cite{Nielsen2000} 

In order to see why WM is useful for our purpose, it is instructive to expand the difference between the entropy and the Markov entropy of a quantum state into manifestly nonnegative quantities. Using the entropy chain rule, $S(\rho)= \sum_{k=1}^{N} S(k | \{<k \})$, with the notation $\{ <k\}=\{1,2,\dots,k-1 \}$, we can obtain the following expression for the difference between the two entropies:
\begin{align}
S_M(\rho)-S(\rho)&=   \sum_{k=1}^{N} S(k | \mathcal{M}_k)_{\rho} - S(k| \{ <k\})_{\rho}. \nonumber
\end{align}
Clearly, the second term cannot be evaluated from the local reduced density matrices in general. We can circumvent this issue by applying WM to a judiciously chosen subsystem:
\begin{equation}
S(k| \{ <k\})_{\rho} + S(k| \mathcal{M}_k')_{\rho} \geq 0, \nonumber
\end{equation}
where $\mathcal{M}_k'$ is a set that is disjoint from $k \cup \{ <k\}$. To summarize, we have the following upper bound on the difference between the two entropies:
\begin{equation}
S_M(\rho) - S(\rho) \leq \sum_{k=1}^{N} S(k | \mathcal{M}_k)_{\rho} + S(k | \mathcal{M}_k')_{\rho}. \label{eq:MarkovSSA_UpperBound}
\end{equation}
It should be noted that $\mathcal{M}_k$ and $\mathcal{M}_k'$ can be chosen arbitrarily, so long as they satisfy the following constraints:
\begin{align}
&\mathcal{M}_k \subset \{<k \} \nonumber \\
&\mathcal{M}_k' \subset \{>k \}, \label{eq:MarkovShieldsConstraints}
\end{align}
where $\{ >k\} = \{k+1, \cdots, N \}$. In particular, $\mathcal{M}_k$ and $\mathcal{M}_k'$ can be chosen to be local, so that the upper bound in Eq.\ref{eq:MarkovSSA_UpperBound} can be evaluated from the local reduced density matrices. Using the fact that $\rho$ and $\sigma$ are consistent over $k \cup \mathcal{M}_k \cup \mathcal{M}_k'$ for all $k$, we obtain the final form of Eq.\ref{eq:Main_Result}:
\begin{equation}
|\rho - \sigma|_1 \leq  2^{3/2}\Big(\sum_{k=1}^{N} S(k|\mathcal{M}_k)_{\rho}+ S(k|\mathcal{M}_k')_{\rho}\Big)^{1/2} \label{eq:Main_Result_Specific}
\end{equation}

{\it Implication of area law---} Now, we apply the area law assumption to Eq.\ref{eq:Main_Result_Specific}, and effectively prove the informational completeness of local observables. More specifically, we focus on a topologically ordered system supported on a 2D square lattice. We begin by coarse-graining the system so that each ``supersites" are sufficiently larger than the correlation length. We shall denote each of these supersites as $k$, and simply refer to them as sites from now on. 

Since the lattice spacing between each sites are much larger than the correlation length, entanglement entropy over a small number of sites can be approximated as follows:
\begin{equation}
S(A) = \alpha l - \gamma + o(1),\label{eq:TEE}
\end{equation}
where $S(A)= -\Tr(\rho_A \log \rho_A)$ is the entanglement entropy of $A$, $l$ is the perimeter of $A$, $\gamma$ is the topological entanglement entropy, and $o(1)$ is the approximation error.\cite{Kitaev2006,Levin2006} 

Our goal is to minimize the upper bound in Eq.\ref{eq:Main_Result_Specific} subject to the three constraints we have discussed so far. First, both $\mathcal{M}_k$ and $\mathcal{M}_k'$ are local. Second, Eq.\ref{eq:MarkovShieldsConstraints} is satisfied. Third, entanglement entropy can be expressed as Eq.\ref{eq:TEE}.
 
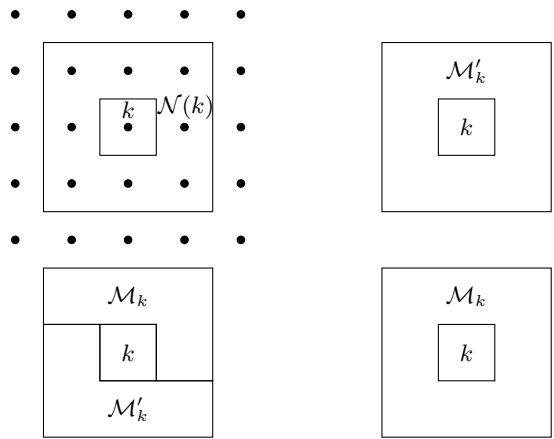
\begin{figure}[h]
\begin{tikzpicture}
\draw[xshift=6*.75cm,yshift=4*.75cm]  (1.5*.75,1.5*.75) rectangle +(3*.75,3*.75);
\draw[xshift=6*.75cm,yshift=4*.75cm]  (2.5*.75,2.5*.75) rectangle +(1*.75,1*.75);
\draw[xshift=6*.75cm,yshift=4*.75cm] (3*.75,3*.75)node {$k$};
\draw[xshift=6*.75cm,yshift=4*.75cm] (3*.75,4*.75)node {$\mathcal{M}_k'$};

\draw[yshift=4*.75cm]  (1.5*.75,1.5*.75) rectangle +(3*.75,3*.75);
\draw[yshift=4*.75cm]  (2.5*.75,2.5*.75) rectangle +(1*.75,1*.75);
\draw[yshift=4*.75cm] (3*.75,3*.75)node[anchor=south] {$k$};
\draw[yshift=4*.75cm] (4.03*.75,3*.75)node[anchor=south] {$\mathcal{N}(k)$};

\draw[] (1.5*.75,1.5*.75) -- ++(0,2*.75) -- ++(1*.75,0)-- ++(0,-1*.75)--++(2*.75,0)--++(0,-1*.75)--cycle;
\draw[] (1.5*.75,3.5*.75)--++(0,1*.75)--++(3*.75,0)--++(0,-2*.75)--++(-1*.75,0)--++(0,1*.75)--cycle;
\draw[]  (2.5*.75,2.5*.75) rectangle +(1*.75,1*.75);
\draw (3*.75,3*.75)node {$k$};
\draw (3*.75,4*.75)node {$\mathcal{M}_k$};
\draw (3*.75,2*.75)node {$\mathcal{M}_k'$};

\draw[xshift=6*.75cm]  (1.5*.75,1.5*.75) rectangle +(3*.75,3*.75);
\draw[xshift=6*.75cm]  (2.5*.75,2.5*.75) rectangle +(1*.75,1*.75);
\draw[xshift=6*.75cm] (3*.75,3*.75)node {$k$};
\draw[xshift=6*.75cm] (3*.75,4*.75)node {$\mathcal{M}_k$};

\foreach \x in {1,...,5}
\foreach \y in {5,...,9}
{
\filldraw[black](\x*.75,\y*.75) circle (.05cm);
}
\end{tikzpicture}
\caption{For a site $k$, $\mathcal{N}(k)$ is a neighbourhood of $k$, a set of sites that are distance $O(1)$ away from $k$. Different choices of $\mathcal{M}_k$ and $\mathcal{M}_k'$ that partition the neighbourhood of $k$ is shown. Under Eq.\ref{eq:TEE}, $S(k | \mathcal{M}_k) + S(k| \mathcal{M}_k')=o(1)$. \label{FIG:LocalConstraints}}
\end{figure}

In order to understand what the optimal choice is, it is instructive to compute $S(k | \mathcal{M}_k) + S(k| \mathcal{M}_k')$ for different choices of $\mathcal{M}_k$ and $\mathcal{M}_k'$. By the virtue of Eq.\ref{eq:TEE}, the result only depends on the topology of the subsystems, up to the $o(1)$ correction. Some examples are described in FIG.\ref{FIG:LocalConstraints}.

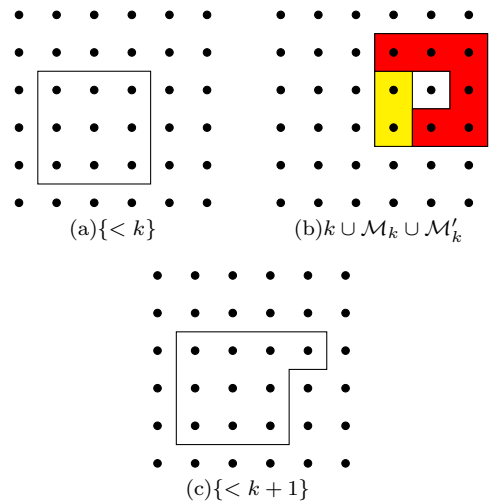
\begin{figure}[h]
	        \subfigure[$\{<k\}$]
	        {
	        \begin{tikzpicture}
	        \foreach \x in {1,...,6}
		\foreach \y in {1,...,6}
			{
				\filldraw[black](\x*.5,\y*.5) circle (.05cm);
			}
		\draw[]  (1.5*.5,1.5*.5) rectangle +(3*.5,3*.5);
		\end{tikzpicture}
	        }\qquad
	        \subfigure[$k \cup \mathcal{M}_k \cup\mathcal{M}_k'$ ]
	        {
	        \begin{tikzpicture}
	        \draw[]  (3.5*.5,2.5*.5) rectangle +(3*.5,3*.5);
		\draw[]  (4.5*.5,3.5*.5) rectangle +(.5,.5);
		\draw[fill=yellow]  (3.5*.5,2.5*.5) rectangle +(.5,2*.5);
		\draw[fill=red] (3.5*.5,4.5*.5)--(3.5*.5,5.5*.5)--(6.5*.5,5.5*.5)--(6.5*.5,2.5*.5)--(4.5*.5,2.5*.5)--(4.5*.5,3.5*.5)--(5.5*.5,3.5*.5)--(5.5*.5,4.5*.5)--cycle;
		\foreach \x in {1,...,6}
		\foreach \y in {1,...,6}
			{
				\filldraw[black](\x*.5,\y*.5) circle (.05cm);
			}
		
		\end{tikzpicture}
	        }
	      \\
	        \subfigure[$\{<k+1\}$ ]
	        {
	        \begin{tikzpicture}
	        \foreach \x in {1,...,6}
		\foreach \y in {1,...,6}
			{
				\filldraw[black](\x*.5,\y*.5) circle (.05cm);
			}
		\draw[]  (1.5*.5,1.5*.5)--(1.5*.5,4.5*.5)--(5.5*.5,4.5*.5)--(5.5*.5,3.5*.5)--(4.5*.5,3.5*.5)--(4.5*.5,1.5*.5)--cycle;
		\end{tikzpicture}
	        }
\caption{(Color online) Suppose we have two states that are equivalent over $\{ <k\}$ and $k\cup \mathcal{M}_k \cup \mathcal{M}_k'$. If $S(k| \mathcal{M}_k) + S(k| \mathcal{M}_k')=o(1)$ for both states, they are approximately equivalent over $\{<k+1\}$. One can recursively use this argument until one cannot find such $\mathcal{M}_k$(yellow) and $\mathcal{M}_k'$(red) anymore. \label{FIG:bound}}	       
\end{figure}

In view of FIG.\ref{FIG:LocalConstraints}, we have a clear strategy; find a sequence of sites that allows a choice of $\mathcal{M}_k$ and $\mathcal{M}_k'$ such that they are topologically equivalent to one of the possibilities depicted in FIG.\ref{FIG:LocalConstraints}. In 2D, it is always possible to find such a sequence as long as $\{<k\}$ are topologically equivalent to each other for all $k$; see FIG.\ref{FIG:bound} for an example. For any operator that is supported on such set, the expectation value of the operator is guaranteed to be consistent.\footnote{Of course, the sum of $o(1)$ terms should be small enough to ensure the bound is nontrivial.}

{\it Estimating and certifying a quantum state---} For systems that attain a nontrivial upper bound on the trace distance, as in FIG.\ref{FIG:bound}, we have two natural applications: quantum state tomography and quantum state verification. More precisely, suppose we have prepared many copies of some multipartite quantum state. We would like to either (i) estimate the state or (ii) verify that it is close to some target state. 

A na\"ive counting on the number of parameters may suggest that one needs to perform exponentially many measurements to estimate a many-body quantum state. Our result shows that this is not quite the case, at least for a large class of physically relevant states. 

In quantum state tomography, one can easily determine whether a given measurement data over the local observables is sufficient to faithfully reconstruct the prepared state. If (i) the local reduced density matrices are measured with a sufficiently high precision and (ii) the upper bound is sufficiently small, the set of local reduced density matrices becomes an accurate representation of the global state. This is due to the fact that one can simply find a global state consistent with the local reduced density matrices; such global state is automatically guaranteed to be close to the prepared state by our bound. Of course, it would be remiss if we do not mention that there is an important caveat. Namely, finding a consistent global state may be computationally difficult. In fact, for a general quantum state, even checking the existence of a consistent global state is QMA-complete.\cite{Liu2007} 

If the objective is to verify whether a prepared state is close to some target state, it is possible to circumvent the potential issue of the computational hardness. One can simply compute the bound from a set of local measurement data. If (i) the bound produces a small number and (ii) the estimated local reduced density matrices are consistent with those of the target state, the prepared state ought to be close to the target state. It will be interesting to compare our protocol to the other efficient state verification protocols that exist in the literature.\cite{Flammia2011,Silva2011}

{\it Equivalence relation between gapped ground states---} Our result shows clearly that a set of local reduced density matrices completely determines the ground state bulk wavefunction of a gapped system, assuming Eq.\ref{eq:TEE} holds. This statement is highly nontrivial in that it is even applicable to states that are long-range entangled, i.e., states with $\gamma>0$.

One may think that it is impossible to faithfully reconstruct a long-range entangled state from purely local information. In light of our result, we believe this intuition has to be modified accordingly. The impossibility only concerns the expectation values of the operators whose support cannot be contracted to a point. All the other expectation values are completely determined by the local reduced density matrices.

Our result also suggests an intriguing equivalence relation that exists between quantum many-body ground states. Namely, equivalence of local reduced density matrices imply the equivalence of the global state under Eq.\ref{eq:TEE}. It will be interesting to understand the consequence of this rather surprising observation, especially in conjunction with the local unitary equivalence framework.\cite{Chen2010}

{\it Discussion---} In order to arrive at our main conclusion, we made three important observations. First, approximately conditionally independent states are essentially defined by their local reduced density matrices. Second, one can certify the smallness of conditional mutual information locally. Third, area law implies the locally checkable upper bound is small. These ideas, combined together, led to a universal certificate that is expected to perform well for a large class of naturally occurring quantum many-body systems. 

Our work opens up a new possibility of efficiently estimating and certifying a highly entangled states of matter that exists in two, and potentially higher dimensions as well. Furthermore, our work clearly shows that a set of local reduced density matrices can be an efficient and accurate representation of the global state even in the presence of long-range entanglement. This suggests an intriguing possibility of systematically studying the properties of such quantum many-body systems from a set of local observables alone.

On a more general ground, we believe the insights obtained in this manuscript will prove useful in other contexts as well. For example, it is well-known that there are difficulties in studying the structure of conditionally independent states that remains to be true for approximately conditionally independent states.\cite{Hayden2004,Ibinson2007} Our result provides an alternative characterization which is weaker than the one discussed in Ref.\cite{Hayden2004}, but has the virtue of being stable. 

Area law is often presented as an evidence that ``physical" states are atypical. Our result corroborates this intuition, with an unexpected twist of surprise: that area law implies informational completeness of local observables.

I am indebted to Alexei Kitaev for inspiration. I also thank Steve Flammia and Olivier Landon-Cardinal for helpful discussions. Research at Perimeter Institute is supported by the Government of Canada through Industry Canada and by the Province of Ontario through the Ministry of Economic Development and Innovation.

\bibliography{bib}

\appendix
\section{Proofs and generalizations}
Suppose we have two multipartite states $\rho$ and $\sigma$ with a uniform upper bound on the trace distance between their local reduced density matrices:
\begin{equation}
|\rho_{k\mathcal{M}_k} -\sigma_{k\mathcal{M}_k}|_1 \leq \epsilon.
\end{equation}

Recently, we proved an inequality that strengthens the concavity of von Neumann entropy\cite{Kim2013a}: 
\begin{equation}
S(c\rho +(1-c)\sigma) - cS(\rho) - (1-c)S(\sigma) \geq \frac{1}{2}c(1-c)|\rho-\sigma|_1^2.
\end{equation}
One can also see \cite{Kim2014} for a simpler proof. Setting $c=\frac{1}{2}$, we have the following sequence of inequalities:
\begin{align}
\frac{1}{4}|\rho-\sigma|_1^2 &\leq 2S(\frac{\rho+\sigma}{2})- S(\rho)-S(\sigma)\\
&\leq 2S_{M}(\frac{\rho+\sigma}{2}) - S(\rho)-S(\sigma),
\end{align}
where we have applied the fact that Markov entropy is an upper bound of the entropy. If $\epsilon=0$, $S_M(\rho)=S_M(\sigma)=S_M(\frac{\rho+\sigma}{2})$, which implies Theorem 1 as well as 2. 

A generalization to the case where $\epsilon\neq 0$ is straightforward. One can use the fact that conditional entropy is continuous.\cite{Alicki2004} This leads to the following additive correction term to Eq.\ref{eq:MEDLowerBound}:
\begin{equation}
\sum_k 4\epsilon \log d_k + 2H(\epsilon),
\end{equation}
where $d_k$ is the dimension of $k$ and $H(x) = -x\log x - (1-x)\log (1-x)$ is the binary entropy.

\end{document}